%% file: main.tex
\documentclass[manuscript,screen,natbib=false]{acmart}

\AtBeginDocument{
  \providecommand\BibTeX{{
    Bib\TeX}}}

\usepackage{amsmath,amsfonts}
\usepackage{algorithmic}
\usepackage{array}
\usepackage[caption=false,font=normalsize,labelfont=sf,textfont=sf]{subfig}
\usepackage{textcomp}
\usepackage{stfloats}
\usepackage{verbatim}
\usepackage{graphicx}
\usepackage[per-mode=symbol,retain-explicit-plus,separate-uncertainty=true,detect-all=true]{siunitx}
\def\defaultDigitGrouping{5}
\def\BibTeX{{\rm B\kern-.05em{\sc i\kern-.025em b}\kern-.08em
    T\kern-.1667em\lower.7ex\hbox{E}\kern-.125emX}}
\usepackage{balance}
\usepackage{biblatex}
\usepackage{caption}
\usepackage{url}
\begin{document}

\title{X-HEEP: An Open-Source, Configurable and Extendible RISC-V Microcontroller for the Exploration of Ultra-Low-Power Edge Accelerators}

\author{Simone~Machetti}
\affiliation{
  \institution{Embedded Systems Laboratory (ESL), EPFL}
  \city{Lausanne}
  \country{Switzerland}}
  \orcid{0000-0002-2887-5031}

\author{Pasquale~Davide~Schiavone}
\affiliation{
  \institution{Embedded Systems Laboratory (ESL), EPFL}
  \city{Lausanne}
  \country{Switzerland}}
  \orcid{0000-0003-2931-0435}

\author{Thomas~Christoph~Müller}
\affiliation{
  \institution{Embedded Systems Laboratory (ESL), EPFL}
  \city{Lausanne}
  \country{Switzerland}}
  \orcid{0009-0004-2805-6310}

\author{Miguel~Pe\'{o}n-Quir\'{o}s}
\affiliation{
  \institution{EcoCloud, EPFL}
  \city{Lausanne}
  \country{Switzerland}}
  \orcid{0000-0002-5760-090X}

\author{David~Atienza}
\affiliation{
  \institution{Embedded Systems Laboratory (ESL), EPFL}
  \city{Lausanne}
  \country{Switzerland}}
  \orcid{0000-0001-9536-4947}

\renewcommand{\shortauthors}{Machetti et al.}
\renewcommand{\shorttitle}{eXtendible Heterogeneous Energy-Efficient Platform (X-HEEP)}

\begin{abstract}

    The field of edge computing has witnessed remarkable growth owing to the increasing demand for real-time processing of data in applications. However, challenges persist due to limitations in the performance and power efficiency of edge-computing devices. To overcome these challenges, heterogeneous architectures have emerged that combine host processors with specialized accelerators tailored to specific applications, leading to improved performance and reduced power consumption. However, most of the existing platforms lack configurability and extendability options, necessitating extensive modifications of the register transfer level (RTL) code for integrating custom accelerators. 

    To overcome these limitations, we introduce in this paper the eXtendible Heterogeneous Energy-Efficient Platform \mbox{(X-HEEP)}. \mbox{X-HEEP} is an open-source platform designed to natively support the integration of ultra-low-power edge accelerators. It provides customization options to match specific application requirements by exploring various core types, bus topologies, and memory addressing modes. It also enables a fine-grained configuration of memory banks to match the constraints of the integrated accelerators. The platform prioritizes energy efficiency by implementing low-power strategies, such as clock-gating and power-gating, and integrating these with connected accelerators through dedicated power control interfaces. 

    We demonstrate the real-world applicability of \mbox{X-HEEP} by providing an integration example tailored for healthcare applications that includes a coarse-grained reconfigurable array (CGRA) and in-memory computing (IMC) accelerators. The resulting design, called \mbox{HEEPocrates}, has been implemented both in field programmable gate arrays (FPGAs) on multiple Xilinx chips, for prototyping and exploration, and in silicon with TSMC \SI{65}{\nano\meter} low-power CMOS technology. The fabricated chip can operate from \SI{0.8}{\volt} to \SI{1.2}{\volt}, achieving a maximum frequency of \SI{170}{\mega\hertz} and \SI{470}{\mega\hertz}, respectively. Its power consumption ranges from \SI{270}{\micro\watt} at \SI{32}{\kilo\hertz} and \SI{0.8}{\volt}, to \SI{48}{\milli\watt} at \SI{470}{\mega\hertz} and \SI{1.2}{\volt}.
    
    We run a set of healthcare applications and measure their energy consumption to demonstrate the alignment of our chip with other state-of-the-art microcontrollers commonly adopted in this domain, showing that \mbox{HEEPocrates} provides a good trade-off between acquisition-dominated and processing-dominated applications for energy efficiency. Moreover, we present the energy benefits of \SI{4.9}{\times} and \SI{4.8}{\times} gained by exploiting the integrated CGRA accelerator and IMC accelerator, respectively, compared to running on the host CPU.
    
\end{abstract}


\maketitle


\input{sections/1-introduction.tex}

\input{sections/2-state-of-the-art.tex}

\input{sections/3-x-heep.tex}

\input{sections/4-heepocrates.tex}

\input{sections/5-experimental-setup.tex}

\input{sections/6-experimental-results.tex}

\input{sections/7-conclusions.tex}

\printbibliography

\end{document}

%% file: sections/1-introduction.tex
\section{Introduction}

    In recent years, the field of edge computing has witnessed remarkable growth and adoption in commercial products. This process has been driven by the increasing demand for real-time computing solutions, particularly Artificial Intelligence (AI) and Machine Learning (ML) algorithms. As data processing at the edge for new edge AI computing has become more prevalent, the performance and power consumption limitations of edge-computing devices have become increasingly apparent, which has posed significant challenges for researchers and engineers.

    Heterogeneous architectures have emerged to overcome these challenges. These architectures offer a promising path toward high energy efficiency while maintaining performance constraints. Heterogeneous architectures rely on a combination of ultra-low-power host processors to run control and communication tasks, and custom accelerators tailored to specific application domains, such as artificial intelligence, image processing, healthcare, and cryptography, to run computationally demanding tasks.

    Control and communication tasks include accessing external memories, acquiring data from analog-to-digital converters (ADCs) or sensors, preparing data for computations, and running real-time operating system (RTOS) functions. Meanwhile, computational demanding tasks focus on data processing, for example, convolutional and fully connected layers in neural networks (NNs), fast-Fourier transforms (FFTs) in temporal series, secure-hash algorithms (SHA) in cryptography, etc.

    Each accelerator comes with unique requirements, such as memory size, area, performance, and power, to meet the constraints of the target applications. For this reason, proper customization of host platforms is imperative. This may include exploring different CPUs to trade performance and power, bus topologies and memory hierarchy, memory sizes to accommodate the required computational data, peripherals to provide the necessary I/O connectivity, power domains and strategies, etc. However, commercial platforms limit hardware exploration due to their non-open-source nature. They often involve costly licensing models and do not allow for customization. As a result, there is a growing preference for open-source platforms as a more attractive solution that does not limit exploration and customization, and that gives designers digital sovereignty and control over IPs.
    
    Today, there are an increasing number of open-source projects related to heterogeneous systems, thanks to the open RISC-V instruction set architecture (ISA) revolution. However, many of such platforms focus only on the CPU part, whereas microcontroller-based state-of-the-art projects lack the flexibility and customization options needed to fulfill accelerator requirements natively. These limitations include restricted configurability for the internal platform's components (core, memory, bus, etc.) to adapt to the application needs, limited support for external accelerator connectivity to communicate with the host system, and inadequate built-in power management strategies to optimize energy efficiency. Thus, hardware developers need to extensively modify the platform to properly align with the target applications on their own copy of the platform. This includes forking, modifying, and maintaining the forked platform's repository, leading to high maintenance costs. Therefore, addressing the configurability and extendability aspects of these platforms is crucial to lowering the adoption barrier of open-source-based edge solutions.
    
    In this paper, we address the limitations mentioned above by introducing \mbox{X-HEEP}\footnote{\textit{X-HEEP} is freely downloadable at https://github.com/esl-epfl/x-heep under a permissive license.} \cite{x-heep}, an open-source configurable and extendable platform designed to support the exploration of ultra-low power edge accelerators. \mbox{X-HEEP} is a streamlined configurable host architecture based on RISC-V and built on top of existing IPs from relevant open-source projects, such as the PULP project, the OpenHW Group, and the OpenTitan project, as extensively verified, validated in silicon, and adopted in commercial products. This allows extensive reuse of third-party hardware and software extensions and inheriting verification and silicon validation. 
    
    To allow users to explore their custom solutions, \mbox{X-HEEP} can be natively extended via the proposed \mbox{eXtendible} \mbox{Accelerator} \mbox{InterFace} (XAIF), which allows the integration of a wide range of accelerators with different area, power, and performance constraints. Having a complete interface that covers all the edge accelerator performance and power requirements will enable extensive reuse of hardware and software IPs, reducing costs and mitigating fragmentation. To explore the custom hardware design space, users will use X-HEEP as an IP which exposes an interface capable of addressing all the edge-computing state-of-the-art requirements for domain-specific applications. To enable a high degree of versatility, such exploration can be performed both on FPGAs or RTL simulators, as well as SystemC for a mixed high-level and RTL simulation environment. Additionally, to offer high degree of optimizations, \mbox{X-HEEP} offers internal configurability options through the selection of different~(1)~core types, depending on the target workloads~\cite{zeroriscy};~(2)~bus topology and addressing mode, ensuring a perfect match with the bandwidth requirements of the integrated accelerators;~(3)~memory size, depending on the processing data and application complexity; and finally~(4)~peripherals, to provide the needed I/O connectivity. This configurability enables designers to tailor the platform to specific application requirements and meet area, power, and performance constraints.
    
    As energy efficiency is a key figure in edge computing devices, \mbox{X-HEEP} implements state-of-the-art fine-grained low-power strategies such as clock-gating, power-gating, and RAM retention, which are integrated into the XAIF interface to be leveraged by the connected accelerators and maximize overall energy efficiency.

    To demonstrate the real-world applicability of \mbox{X-HEEP}, we present an integration example specifically tailored for ultra-low-power healthcare applications. These applications typically involve long and slow acquisition periods, where data from external bio-sensors are stored in memory while the rest of the system is in an idle state, followed by intense processing periods to compute pattern extraction algorithms based on digital signal processing algorithms, or machine learning (ML), and deep learning. Therefore, we extended \mbox{X-HEEP} with a CGRA accelerator~\cite{cgra} and an IMC accelerator~\cite{blade}, both of which have been shown to efficiently reduce the overall energy consumption of healthcare applications~\cite{ieeemicro,ponzinamachettimicro}. We configured \mbox{X-HEEP} with the RISC-V OpenHW Group CV32E20 core~\cite{zeroriscy}, 8 banks of \SI{32}{\kibi\byte} on-chip SRAM organized in a contiguous addressing mode, and 11 different power domains (including the external accelerators) that can be individually switched on and off for fine-grained power control. 
    
    The resulting design, called \mbox{HEEPocrates}, has been implemented both in FPGAs on the Zynq 7020, Zynq UltraScale+, and Artix 7 chips by Xilinx, for early prototyping, verification, and system exploration, as well as in silicon with TSMC \SI{65}{\nano\meter} low-power CMOS technology, for silicon validation and profiling performance, power, and area figures. The measured performance of the fabricated chip shows that it can operate in a wide voltage range, from \SI{0.8}{V} to \SI{1.2}{V}, achieving a maximum frequency of \SI{170}{\mega\hertz} and \SI{470}{\mega\hertz}, respectively. Its power consumption ranges from \SI{270}{\micro\watt} at \SI{32}{\kilo\hertz} and \SI{0.8}{\volt}, to \SI{48}{\milli\watt} at \SI{470}{\mega\hertz} and \SI{1.2}{\volt}.

    To validate our design and compare it with state-of-the-art solutions, we measured the energy consumption of the chip in a set of healthcare applications, showing that it offers a good trade-off between the computationally hungry and acquisition-dominated state-of-the-art microcontrollers commonly adopted in this domain. This demonstrates the flexibility of our host platform in adapting to the specific needs of integrated accelerators and matching the strict requirements of healthcare applications. Furthermore, we present the energy benefits of \SI{4.9}{\times} and \SI{4.8}{\times} gained by exploiting the integrated CGRA accelerator and the IMC accelerator, respectively, compared to running on the host CPU.

    Throughout this work, we will dive deeper into the features of the \mbox{X-HEEP} platform, describing its architecture, configurability, and extendability options to build versatile and energy-efficient edge applications.

    The following contributions are presented:

    \begin{itemize}
        \item \mbox{X-HEEP}: A configurable and extendible RISC-V microcontroller to support the exploration of ultra-low-power edge accelerators.
        \item \mbox{XAIF}: A configurable interface that adapts to the different requirements of accelerators in terms of programmability, bandwidth, interrupts, and power modes, and that allows their seamless integration into the \mbox{X-HEEP} architecture.
        \item \mbox{HEEPocrates}: A real-world integration example, based on TSMC \SI{65}{\nano\meter} low-power CMOS technology, that includes a CGRA and an IMC accelerator.
        \item An open-source repository with a permissive license with the complete \mbox{X-HEEP} platform code and documentation to allow researchers to explore new custom accelerators and advance research in this field. 
    \end{itemize}

    The remainder of this paper is structured as follows. In Section~2, we conduct an in-depth analysis of the most relevant state-of-the-art accelerators and host platforms. In Section~3, we provide a qualitative and quantitative description of the configurability and extendability features of \mbox{X-HEEP}. In Section~4, we present a real-world integration example, called \mbox{HEEPocrates}. In Section~5 we describe our experimental setup, while in Section~6 our experimental results. Lastly, in Section~7, we offer a comprehensive summary of the main conclusions of our work.

%% file: sections/2-state-of-the-art.tex
\section{State-of-the-art}

    This section gives an overview of cutting-edge accelerators, analyzing the fundamental requirements necessary for their integration into host platforms. Such requirements are collected into the XAIF to accommodate all the state-of-the-art accelerators. Subsequently, it focuses on host platforms from the state of the art, conducting an evaluation of their strengths and limitations in terms of configurability and extendability.

    \subsection{Edge-computing accelerators}  \label{sec:accel}

        The extensive array of open-source accelerators includes a diverse range of requirements regarding memory capacity, area, performance, and power efficiency. Therefore, an analysis of their main features becomes imperative for the design of flexible and efficient host platforms.
        
        We can divide accelerators into three main categories: memories, processors (and co-processors), and I/O peripherals. One accelerator can belong to one or more categories.
        
        \subsubsection{Memories} \label{sec:mem}
        
            Memory accelerators are a class of IPs that feature one or more slave ports to access internal functionality. These IPs require the host CPU, or a DMA, to copy the needed computational data from the main memory of the platform to their internal data memory, before starting the operations. At the end of the computations, an interrupt or status bit could be used to synchronize with the host CPU.

            An example is the Keccak accelerator presented in~\cite{keccak}. This accelerator exposes two \SI{32}{\bit} slave ports, one to access the internal register file used for control and status operations and one to access the private data memory, which stores the processing data.

            Other examples are in-memory or near-memory macros such as the C-SRAM~\cite{cea_imc}, where the IP is connected via a \SI{32}{\bit} slave port to a host platform that sends commands/instructions to the memory through write operations. C-SRAM decodes the memory instructions, by concatenating the address and the write data transmitted by the CPU, and performs the requested operation.

        \subsubsection{Processors} \label{sec:proc}
        
            To improve performance, many accelerators feature one or multiple master ports to independently read in parallel the processing data and write back the generated results from/to the main memory. 
            
            Some examples are domain-specific accelerators, such as DSP engines~\cite{dsp_coproc}, CGRAs~\cite{cgra}, multi-CPU clusters~\cite{pulp_cluster}, GPUs~\cite{vortex}, etc., and application-specific accelerators for neural networks~\cite{cnn-accel}, FFT~\cite{fft}, cryptography~\cite{keccak}, image processing~\cite{image}, etc.

            An example of a domain-specific accelerator is the CGRA presented in~\cite{cgra}, which has two \SI{32}{\bit} slave ports for configuration registers and private instruction memory and four \SI{32}{\bit} master ports for reading and writing data from and to the main memory, reaching a maximum bandwidth of \SI{128}{\bit} per bus cycle.
            
            Another example is the PULP cluster~\cite{pulp_cluster}, which features four to eight CV32E40P cores~\cite{ri5cy} connected to a shared instruction cache and a multi-bank scratchpad memory. The cluster exposes one \SI{32}{\bit} slave port for configuration and for pre-loading the memories, and one \SI{64}{\bit} master port shared between the cluster DMA, to transfer data in and out of the scratchpad memory, and the instruction cache, to fetch program code.
        
            Examples of application-specific accelerators are Echoes~\cite{fft} and Marsellus~\cite{cnn}. The former is used to speed up FFT execution and has eight \SI{32}{\bit} master ports, four allocated for input and four for output, reaching a maximum bandwidth of \SI{256}{\bit} per bus cycle. The latter accelerates convolution layers and offers nine \SI{32}{\bit} master ports, with a maximum bandwidth of \SI{288}{\bit} per bus cycle.
        
            Co-processors are a sub-category of processors used to implement custom ISA extensions. Co-processors are either tightly coupled in the processor pipeline, or integrated via a dedicated interface for reusability. 
            
            Their wide application domains include floating-point operations~\cite{fp}, posit arithmetic~\cite{posit}, post-quantum cryptography~\cite{pqcry}, integer complex arithmetic~\cite{complex}, etc. For example,~\cite{pqcry} proposes a post-quantum cryptography ISA extension and interacts with the coupled RISC-V CPU thanks to the CORE-V-XIF \cite{core-v-xif} interface of OpenHW Group.

            All these previous examples illustrate that there is a large choice of possible processors and co-processors for edge AI systems today. Therefore, it is required to have a fast and scalable exploration and prototyping framework to choose the right set of components, co-processors, or domain-specific accelerators that a final implementation should have, and then a well-tuned silicon design flow with a predefined set of open-source hardware components and peripherals.

        \subsubsection{I/O peripherals} \label{sec:io}
        
            These IPs are meant to implement special interfaces to communicate with off-chip components or to pre/post-process data during such communications. 
            
            An example can be found in Arnold~\cite{arnold}, where the embedded FPGA (eFPGA) can be used to control an off-chip accelerator, which requires a custom interface, as well as to pre-process data coming from peripherals before being stored in memory.
            
            In this case, there is a clear need to ideally have a framework that can target both on-chip and off-chip accelerator concepts by enabling a flexible set of interconnect standards. This set of standards should be extendable with minimum effort from the system designer thanks to an interface that enables a superset of interconnection protocols, as we propose in X-HEEP. 
    
    \subsection{Host platforms}

        In this subsection, we present a comparison of relevant open-source platforms that can be used to host edge-computing accelerators, focusing on configurability, extendability, and other key features. The limitations of each platform are analyzed in detail to motivate the need for a dedicated solution that can fulfill all the requirements.

        \subsubsection{PULPissimo~\cite{pulpissimo}}
        
            A single-core platform within the PULP family, designed to target ultra-low-power edge-computing applications. Depending on performance requirements, designers can configure the platform with the CV32E20~\cite{zeroriscy} or CV32E40P~\cite{ri5cy} cores. PULPissimo has been integrated with various accelerators, including the aforementioned multi-CPU cluster~\cite{pulp_cluster}, neural network accelerators~\cite{cnn}, CGRAs~\cite{cgra}, eFPGAs~\cite{arnold}, etc. Many silicon prototypes have been implemented, which demonstrate best-in-class energy efficiency in a wide range of applications.
        
            However, PULPissimo provides only a generic AXI \SI{32}{\bit} slave and a \SI{64}{\bit} master external interfaces that are used to connect the multi-CPU cluster, while the other accelerators have been integrated by forking and modifying the original RTL code. Such external interfaces may limit accelerators' bandwidth. Moreover, the platform lacks native support for external interrupts/events and power control, which is crucial for efficient power management. Lastly, the platform does not offer configurability options to select memory size, bus topology, and memory addressing mode, or to change the included peripherals, which limit area, bandwidth, and power-space exploration.

        \subsubsection{Cheshire~\cite{cheshire}}
        
            The limitations mentioned above have been partially addressed by another PULP-based platform, called Cheshire. Cheshire is based on the CVA6 core~\cite{ariane} and allows designers to choose the number of external slave and master ports to connect their custom accelerators. Furthermore, the platform allows for the configuration of the internal last-level cache (LLC) size and of the necessary peripherals, providing the flexibility needed to target specific application requirements.
        
            However, Cheshire has been designed for high-performance systems and consumes up to \SI{300}{\milli\watt}, making it unsuitable for most ultra-low-power devices, which typically operate in the range of tens of \si{\milli\watt}. Furthermore, Cheshire lacks support for external interrupts and power control, which has implications for its overall energy efficiency, as the accelerators are usually power-hungry. Lastly, designers do not have the option to select the core type, bus topology, and memory addressing mode.

        \subsubsection{BlackParrot~\cite{blackparrot}}
        
            An open-source Linux-capable platform designed to accommodate one or multiple custom-designed accelerators. The platform showcases a mesh of heterogeneous tiles, offering the flexibility to compose \SI{64}{\bit} BlackParrot cores, L2 cache slices, I/O, DRAM controllers, and accelerators in various configurations.
        
            However, it does not allow for selecting the core type, bus topology, and memory addressing mode. Additionally, the absence of essential peripherals commonly used in edge devices, such as I2Cs, GPIOs, timers, DMAs, interrupt controllers, and a power manager to implement low-power strategies, restricts the usage of the platform for real applications deployed on ultra-low-power edge applications. Moreover, the platform's internal integration of accelerators, as opposed to external plug-ins, involves forking and modifying the original RTL code, leading to greater effort and higher development costs. Lastly, the \SI{64}{\bit} architecture of BlackParrot targets high-performance systems and is unsuitable for ultra-low-power edge devices.

        \subsubsection{OpenTitan~\cite{opentitan}}
        
            OpenTitan is designed for ultra-low-power edge-secure applications. It offers a single-core architecture based on the CV32E20~\cite{zeroriscy} core and an extensive portfolio of peripherals.
        
            Despite these strengths, OpenTitan does not offer external support for accelerator plug-ins, requiring designers to manually modify the RTL code to integrate their custom accelerators. Furthermore, the platform lacks configurability for core type, bus topology, and memory addressing mode and size. Furthermore, OpenTitan does not come equipped with built-in low-power strategies.

        \subsubsection{Chipyard~\cite{chipyard}}
        
            On the contrary, the Rocket chip generator~\cite{rocket}, which has been subsequently incorporated and expanded into the Chipyard platform, offers extensive configuration options. Using the open-source Chisel hardware description language, designers can craft their system, providing flexibility and customization. The platform offers a wide range of core types, including Ariane, CV32E20, Rocket, and BOOM, allowing designers to tailor the system's performance to meet specific application requirements. Additionally, the memory size and peripherals can be customized, further enhancing its adaptability.
        
            However, even though Chipyard enables accelerators to be integrated into the design using the Chisel language, the platform does not offer external master and slave ports for the connectivity of accelerators. As a result, designers need to invest time in becoming familiar with the Chisel language to successfully configure the architecture and integrate custom accelerators. Furthermore, Chipyard does not provide support for any specific power reduction strategies. Given the critical importance of power efficiency in ultra-low-power applications, designers are forced to implement power-saving techniques manually to achieve the desired energy efficiency level.

        \subsubsection{LiteX~\cite{litex} and ESP~\cite{esp}}
        
            Two other notable SoC generators are LiteX and ESP. LiteX serves as a framework thought to explore various FPGA-based architectures. On the other hand, ESP is an open-source platform designed for heterogeneous SoC design and prototyping on FPGAs. Both platforms offer configurable options, allowing designers to customize core type, memory size, peripherals, and the number of external master and slave ports, making them adaptable to various application requirements.
        
            However, LiteX and ESP focus on FPGA development only and do not offer support for ASIC design flow. Such limitations hinder their applicability in projects aimed at silicon implementations and present difficulties in accurately estimating the platform energy consumption, crucial when evaluating the impact of integrated accelerators. Moreover, they lack built-in support for external interrupts and power control, essential for efficient power management.

        \subsubsection{X-HEEP}

            To overcome the limitations mentioned above and cater to the unique needs of ultra-low-power edge designers, we present in this paper the \mbox{X-HEEP} platform. The proposed platform features a streamlined architecture that operates in conjunction with dedicated open-source tools. These tools enable developers to easily customize and extend the architecture with their accelerators and interconnection interfaces, thus eliminating the need for manual modification of the RTL code. Using \mbox{X-HEEP}, designers can achieve the desired level of configurability, extendability, and power efficiency, making it an ideal choice for a wide range of ultra-low-power edge applications. In addition, it has been developed using SystemVerilog, to offer high compatibility with most of the available electronic design automation (EDA) tools.

%% file: sections/3-x-heep.tex
\section{X-HEEP}

    In this section, we present a qualitative and quantitative analysis of the key features of \mbox{X-HEEP} regarding configurability, extendability, and software support. We synthesized \mbox{X-HEEP} with TSMC \SI{65}{\nano\meter} low-power technology and performed our quantitative analysis at the nominal voltage, \SI{1.2}{\volt}. 

    \subsection{Architecture}

        Figure \ref{fig:x-heep} shows the \mbox{X-HEEP} architecture and its essential components. These include a configurable RISC-V CPU, a configurable bus, a configurable memory, two configurable peripheral domains, and a debug unit. 
        
        \mbox{X-HEEP} leverages existing widely adopted open-source IPs to maintain compatibility with existing systems and reuse available software routines and hardware extensions. Among the wide portfolio of open-source IPs, we selected those that provide permissive licenses, to ease the \mbox{X-HEEP} adoption to a wide range of users, and written in SystemVerilog, to make the integration in existing systems and EDA tools compatible with industrial standards.
        
        The RISC-V cores have been selected from the OpenHW Group CORE-V family, as extensively verified, mature, and implemented in silicon many times; the bus, the memory models, the debug unit, and a plethora of IPs from the PULP project, as again adopted by several stakeholders and validated in silicon multiple times; and the peripherals from the OpenTitan project as documented, verified, and inclusive of hardware-abstraction-layer (HAL) functions. Moreover, \mbox{X-HEEP} includes home-made IPs such as a boot ROM, a power manager, a fast interrupt controller, and a DMA.

        \subsubsection{CPU}

            The user can choose among the CV32E20, CV32E40X, and CV32E40P as core options~\cite{zeroriscy}, to trade off power and performance. In particular, the CV32E20 core is optimized for control-oriented tasks, while the CV32E40P core is optimized for processing-oriented tasks. The CV32E40X core offers power consumption and performance similar to the CV32E40P core, without featuring the floating-point RVF and custom Xpulp ISA extensions. Moreover, it provides an external interface, known as CORE-V-XIF~\cite{core-v-xif}, that allows for the plug-in of custom co-processors to extend the RISC-V ISA without the need to modify the RTL code of the core.

        \subsubsection{Memory}

            The user can select the memory size and number of memory banks to trade off area, power, and storage capacity. Each bank offers a retention state aimed at reducing leakage power, of about \SI{42.5}{\percent} compared to active leakage, when the bank is not accessed for some time but the data needs to be preserved.
            
        \subsubsection{Bus}

            To maximize compatibility with the other IPs selected from the OpenHW Group, PULP platforms, and OpenTitan project, the bus is based on the same open-bus interface (OBI)~\cite{obi} protocol.

            The user can choose either a one-at-a-time topology, where only one master at a time can access the bus (one decoder), or a fully connected topology (same number of decoders as simultaneous masters), where multiple masters can access multiple slaves in parallel, to trade off area and bandwidth. When the fully connected option is used, the user can further configure the bus to access a variable number of banks in a contiguous or interleaved addressing mode. The contiguous mode offers limited bandwidth to applications that require multiple masters to access contiguous data stored in memory but allows for power-gating or setting in retention mode the banks that are not actively used. Vice versa, the interleaved mode offers higher bandwidth to applications that access contiguous data in memory, at the cost of keeping all the banks active all the time. 
            
            In addition, to connect additional components, the bus also exposes a configurable number of slave and master ports to the external XAIF interface to accommodate one or multiple accelerators with different bandwidth constraints.

\begin{figure*} [t]
                \centering
                \captionsetup{justification=centering}
                \subfloat[Bus: one-at-a-time.]{\includegraphics[width = 2.9in]{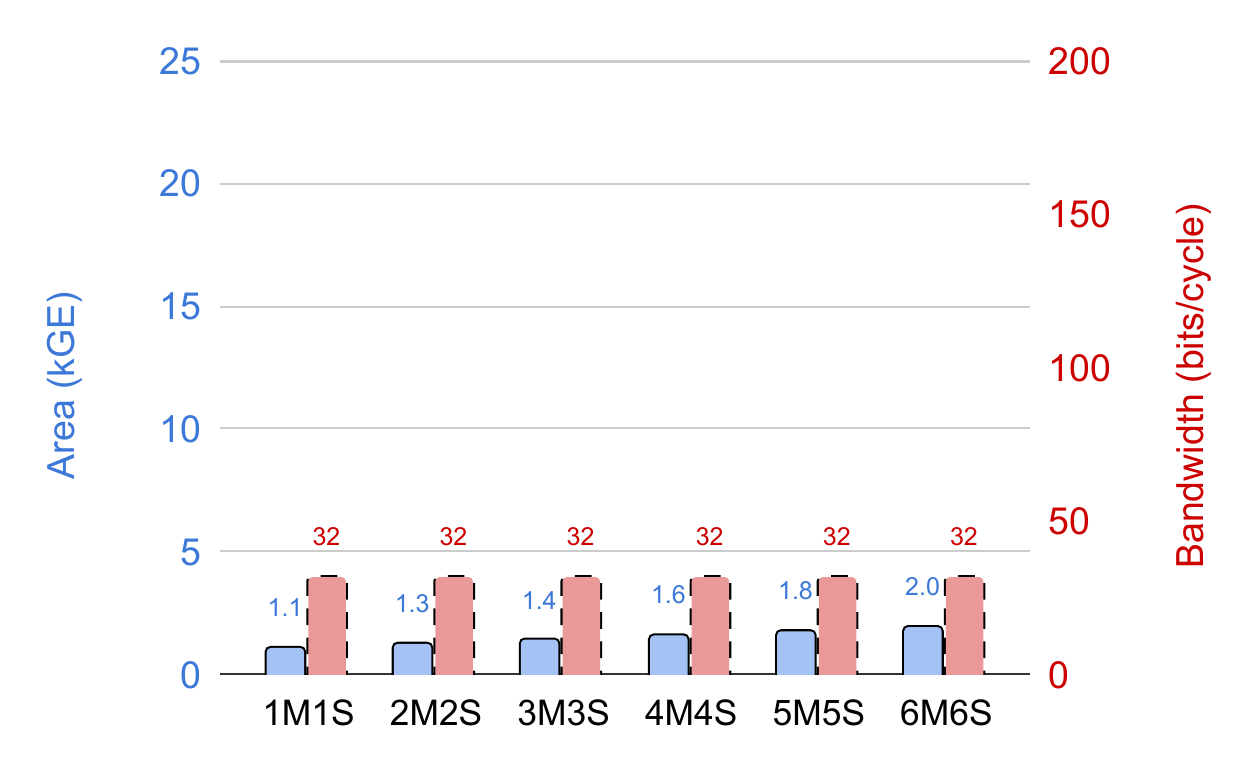}} 
                \subfloat[Bus: fully connected.]{\includegraphics[width = 2.9in]{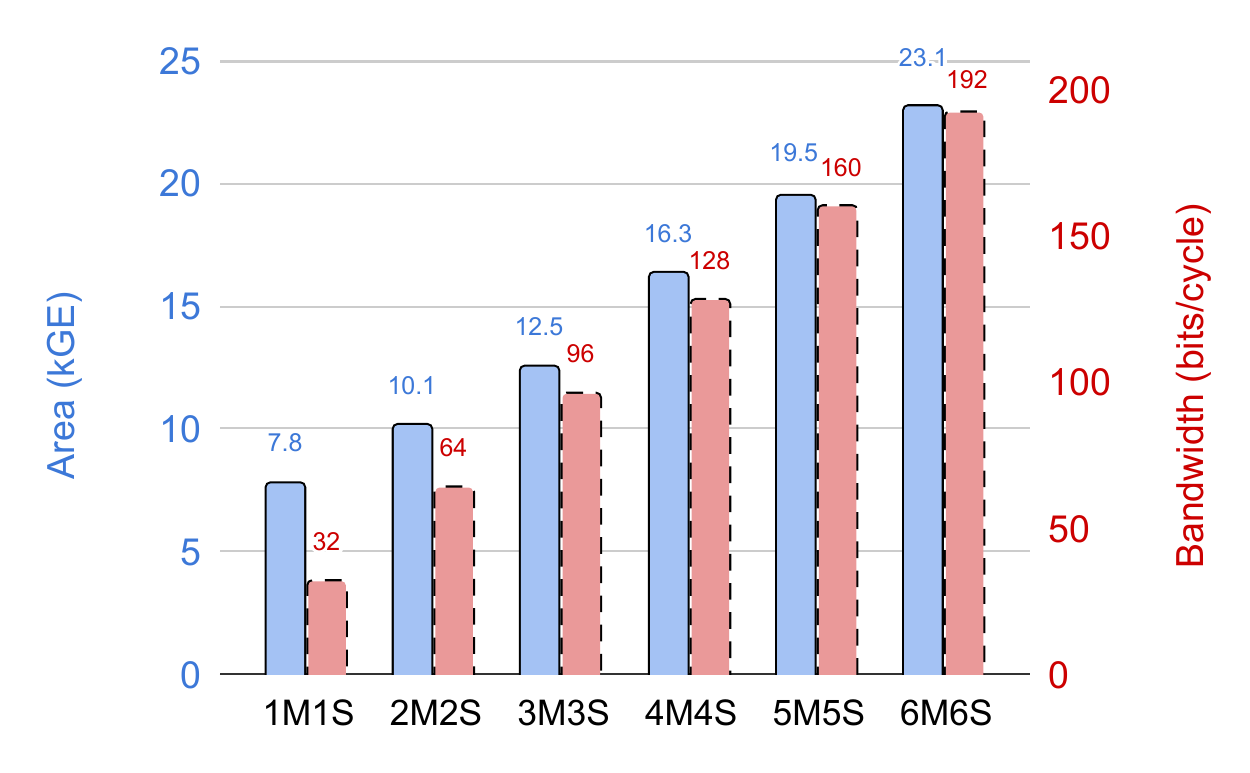}}\\
                \subfloat[Peripherals.]{\includegraphics[width=2.9in]{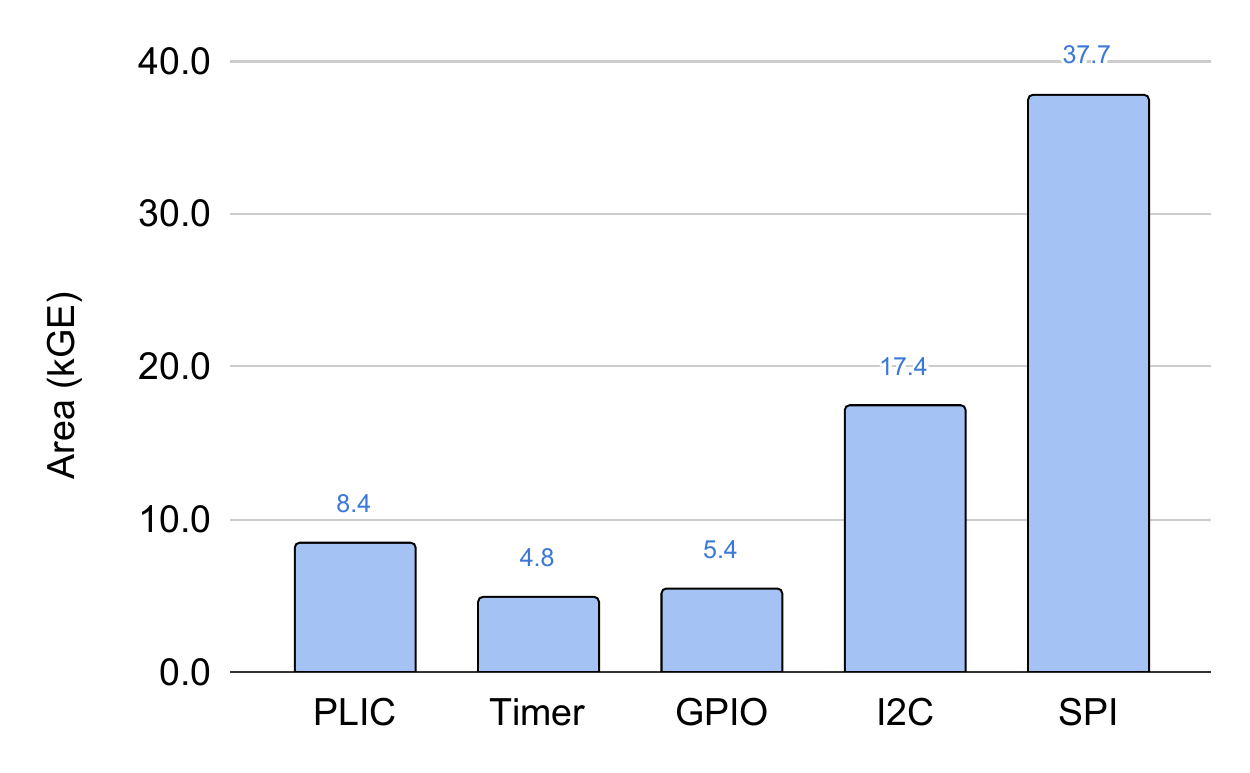}} 
                \subfloat[Power domains.]{\includegraphics[width=2.9in]{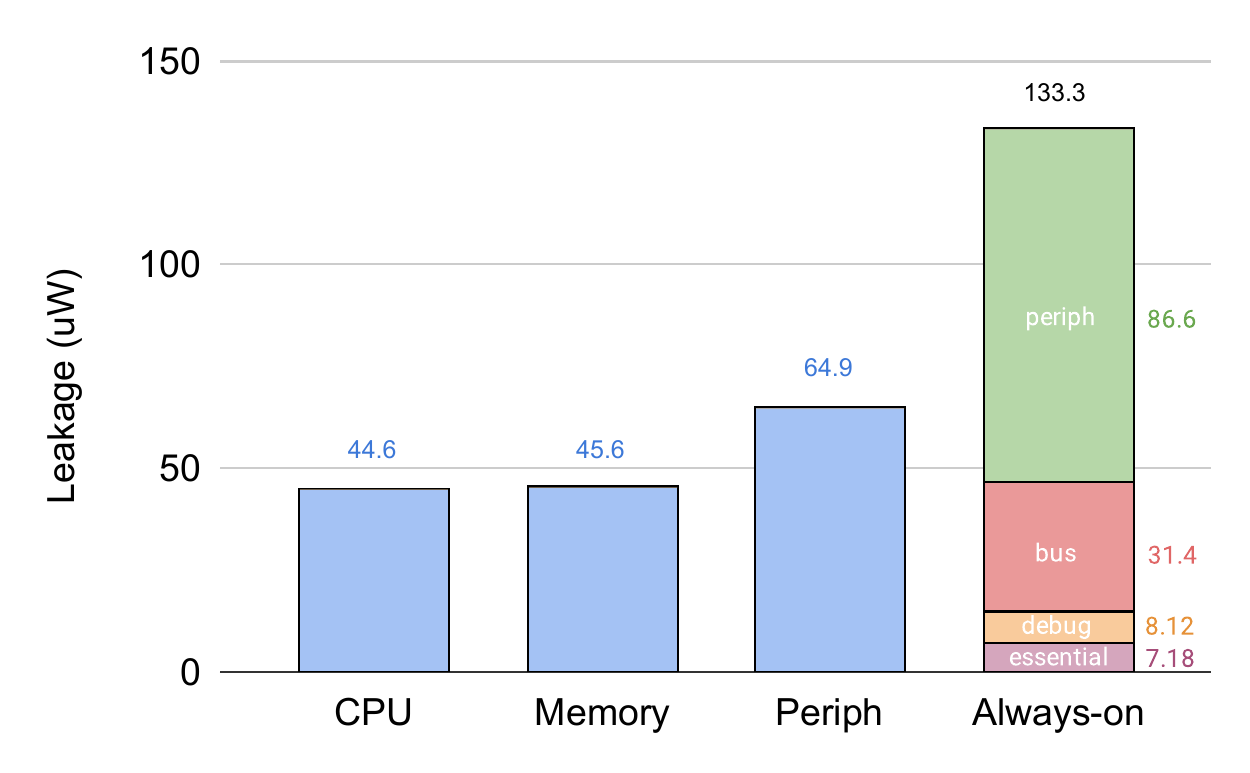}}
                \vspace{5px}
                \caption{Exploration of different \mbox{X-HEEP} configurations. The used technology is TSMC \SI{65}{\nano\meter} low-power CMOS at \SI{1.2}{\volt}.}
                \label{fig:exploration}
            \end{figure*}

            Figures \ref{fig:exploration} (a) and (b) show the variation in the area and the bandwidth of the \mbox{X-HEEP} bus by adding slave/master ports to the basic bus configuration, which connects the CV32E20 core, two memory banks, the debug unit, the two peripheral domains, and no external connection. Ports are added in pairs, i.e., for each external master port (M), we add an internal slave port (S) for a memory bank to avoid limiting bandwidth during memory access. 
            
            Increasing the number of slave/master ports does not lead to any performance improvement in the one-at-a-time configuration, limited to \SI{32}{\bit} per bus cycle according to its architecture. On the contrary, the fully connected configuration maximizes bandwidth, which increases linearly with the number of bus ports, at the cost of a higher area (and power consumption). The bus in the one-at-a-time configuration occupies about \SI{85}{\percent} less silicon space compared to the fully connected configuration, considering the same number of slave/master ports.

            In overall performance, a $16 \times 16$ matrix multiplication algorithm on \mbox{X-HEEP} takes approximately \SI{34}{\percent} fewer clock cycles in the highest performance configuration with the CV32E40P core and fully connected bus compared to the lowest power configuration with the CV32E40P core and one-at-a-time bus. Furthermore, when using the Xpulp extensions and fully connected bus, the CV32E40P can compute matrix multiplication algorithms \SI{4}{\times} faster with \SI{32}{\bit} data or up to \SI{16}{\times} faster with \SI{8}{\bit} SIMD extensions for the same CPU without extensions, as shown in \cite{ri5cy}.

            \begin{figure*} [t]
                \centering
                \includegraphics[width=0.9\textwidth]{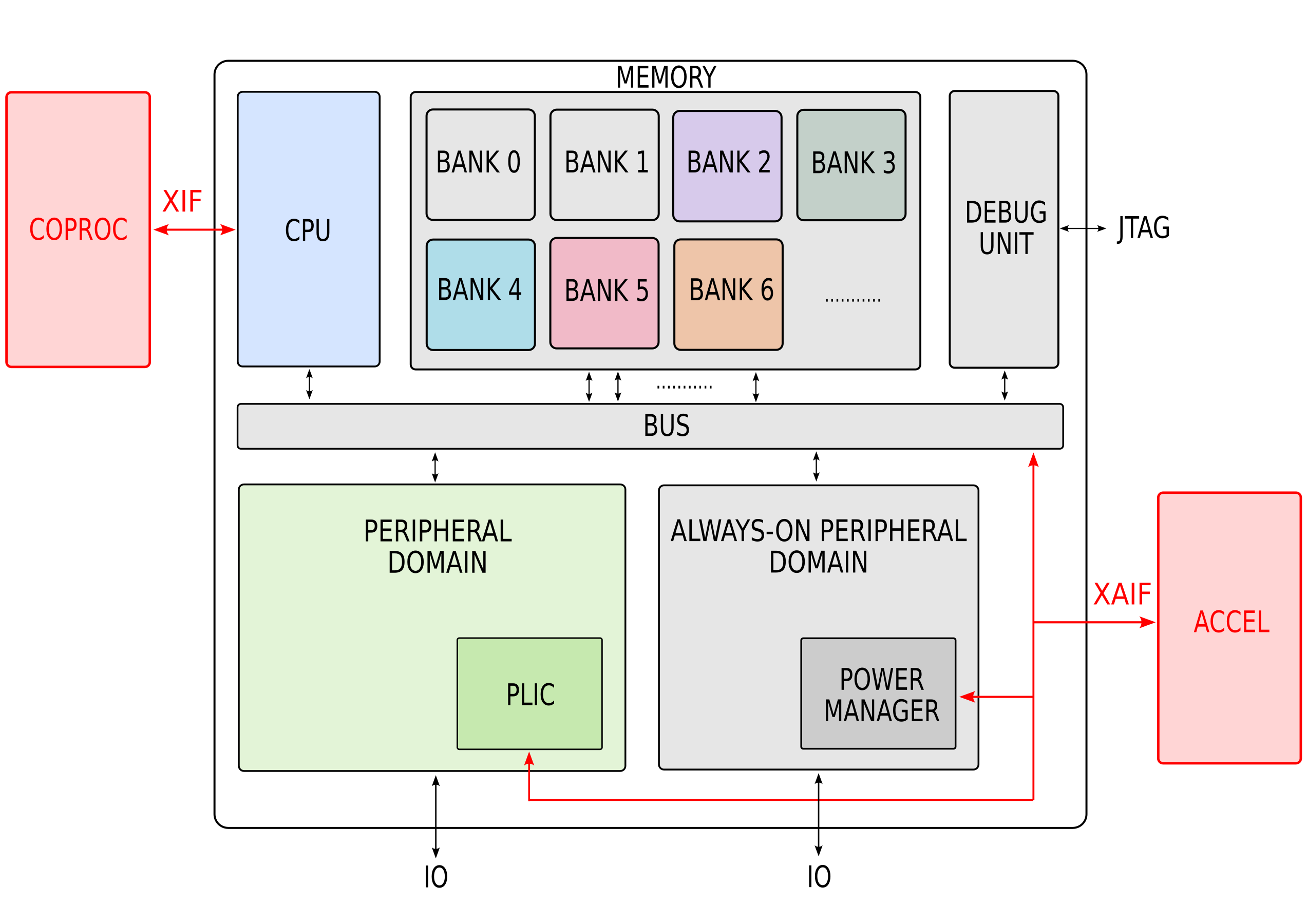}
                \caption{\textit{X-HEEP} architecture. The various power domains are visually marked using different colors. The components in grey are always on. The accelerator and co-processor integration are highlighted in red.}
                \label{fig:x-heep}
            \end{figure*}

        \subsubsection{Peripheral domain}

            Figure \ref{fig:exploration} (c) shows the area of the IPs located in the peripheral domain. This domain includes peripherals that can be removed from the design or powered off if not needed to trade off area or power and functionality. These include a platform-level interrupt controller (PLIC), a timer, and general-purpose I/O peripherals such as a GPIO, I2C, and SPI.

        \subsubsection{Always-on peripheral domain}

            This domain includes IPs that are always powered on. To meet our specific needs and requirements, we custom-designed key components such as an SoC controller, a boot ROM, a power manager, a fast interrupt controller, and a DMA. The domain also includes other peripherals such as a timer, a UART, an SPI, and a GPIO.
    
            The power manager is responsible for implementing low-power strategies, including clock-gating, power-gating, and RAM retention. It features a set of configuration registers that provide the user with real-time control over the available low-power techniques.
            
            The architecture is divided into several power domains, marked with different colors in Figure \ref{fig:x-heep}. Clock-gating can be applied to the main CPU, peripheral domain, and each memory bank, while retention can only be applied to memory banks. Additionally, each power domain can be individually power-gated. The leakage power consumption of each domain is reported in Figure \ref{fig:exploration} (d). The system bus, debug unit, and other essential IPs represent about \SI{35}{\percent} of the leakage power of the always-on domain. The remaining \SI{65}{\percent} comes from other general-purpose peripherals added to enhance versatility, such as a GPIO, SPI, UART, etc.

            The platform can be extended with additional power domains to include user external accelerators. This is possible thanks to external power ports, part of the XAIF interface, directly connected to the power manager, which can be used to clock-gate, power-gate, or set in retention mode external accelerators.

        \subsection{Extendible Accelerator InterFace (XAIF)}
        
            The extensive array of domain-specific hardware accelerators encompasses a diverse range of requirements, including memory capacity, area, performance, and power efficiency. These varied demands are aggregated into the configurable XAIF interface, facilitating enhanced connectivity to state-of-the-art accelerators, and agile integration into microcontrollers for real-life applications. Such an interface gathers all the requirements to extend X-HEEP with domain-specific customizations. To the best of the authors' knowledge, no other open-source platform for edge-computing applications exists that provides such a complete extension interface to fulfill the requirements of state-of-the-art solutions.
        
            \subsubsection{Memory mapped ports}
            
                A configurable number of slave and master ports, utilizing the OBI protocol, can be harnessed to connect custom accelerators to the \mbox{X-HEEP} bus. Slave ports provide easy access and configuration for memory-like accelerators, exemplified in Subsection~\ref{sec:mem}, such as the Keccak~\cite{keccak}, which requires two \SI{32}{\bit} slave ports for control and status operations and data memory. In addition, a further peripheral interface connected to the X-HEEP peripheral bus is provided for external custom peripherals. This peripheral interface is further extended by a FIFO interface to allow easy DMA-peripheral connections. This allows the CPU to wait for peripheral transactions to transfer all data to the main memory with the support of the system DMA, as implemented in~\cite{arnold}. On the other hand, master ports accommodate the bandwidth requirements of processor-like accelerators outlined in Subsection~\ref{sec:proc}. For example, a CGRA~\cite{cgra} leverages the four \SI{32}{\bit} master ports that are used to independently read and write data to and from the main memory.

            \subsubsection{Interrupt ports}
            
                A configurable number of interrupt lines can be used by the custom hardware to rapidly synchronize with the host CPU. Each line is connected to the \mbox{X-HEEP} PLIC interrupt controller, which can be controlled via software. This functionality allows the host CPU to enter a sleep state during active accelerator periods, significantly reducing the overall energy consumption of the running application.
        
            \subsubsection{Power control ports}
            
                To provide low-power strategy capabilities to custom accelerators, a configurable number of power control interfaces is provided. Each interface is connected to the \mbox{X-HEEP} power manager to implement different power-saving strategies. Each interface includes control signals for power-gating, clock-gating, and RAM retention.

    \subsection{Tools and software}

        We present the tools and software provided by \mbox{X-HEEP} to configure, program, and implement user designs.

        \subsubsection{Configuration}
        
            \mbox{X-HEEP} is configured through SystemVerilog templates, which function as a dynamic tool that enables users to automatically customize the RTL code of the platform thanks to customizable parameters. This makes the generated code readable and easy to maintain and debug.

        \subsubsection{Software}
        
            \mbox{X-HEEP} includes a HAL to access peripheral functionalities and supports FreeRTOS for improved development and efficient resource management.
    
        \subsubsection{Simulation and implementation}
        
            \mbox{X-HEEP} offers support for simulation and implementation, in FPGA and silicon, based on the FuseSoc build system~\cite{fusesoc}. FuseSoC supports several EDA tools, such as Verilator, Questasim, Design Compiler, Genus, and Vivado, and automatically generates the scripts required to simulate or implement user designs. Thanks to it, the user can explore the design both at a high level by integrating accelerators described in SystemC, at the RTL level, and FPGA, for early prototyping and exploration, as well as in silicon, for final validation.

%% file: sections/4-heepocrates.tex
\section{HEEPocrates}

    In this section, we present an integration example to demonstrate the real-world applicability of \mbox{X-HEEP}. This integration effort results in \mbox{HEEPocrates}, a heterogeneous architecture designed for ultra-low-power healthcare applications. These applications typically involve extended data acquisition periods during which data from external biosensors are stored in memory, followed by intensive processing periods to classify such data. Therefore, we exploited the XAIF interface to extend \mbox{X-HEEP} with a CGRA accelerator~\cite{cgra} and an IMC accelerator~\cite{blade}, both of which have been shown to efficiently reduce the overall energy consumption of healthcare applications~\cite{ieeemicro,ponzinamachettimicro}. Moreover, each accelerator is located in a separate power domain that can be individually switched on and off for fine-grained power control.

    \subsection{Architecture}

\begin{figure*}
                \centering
                \includegraphics[width=0.7\textwidth]{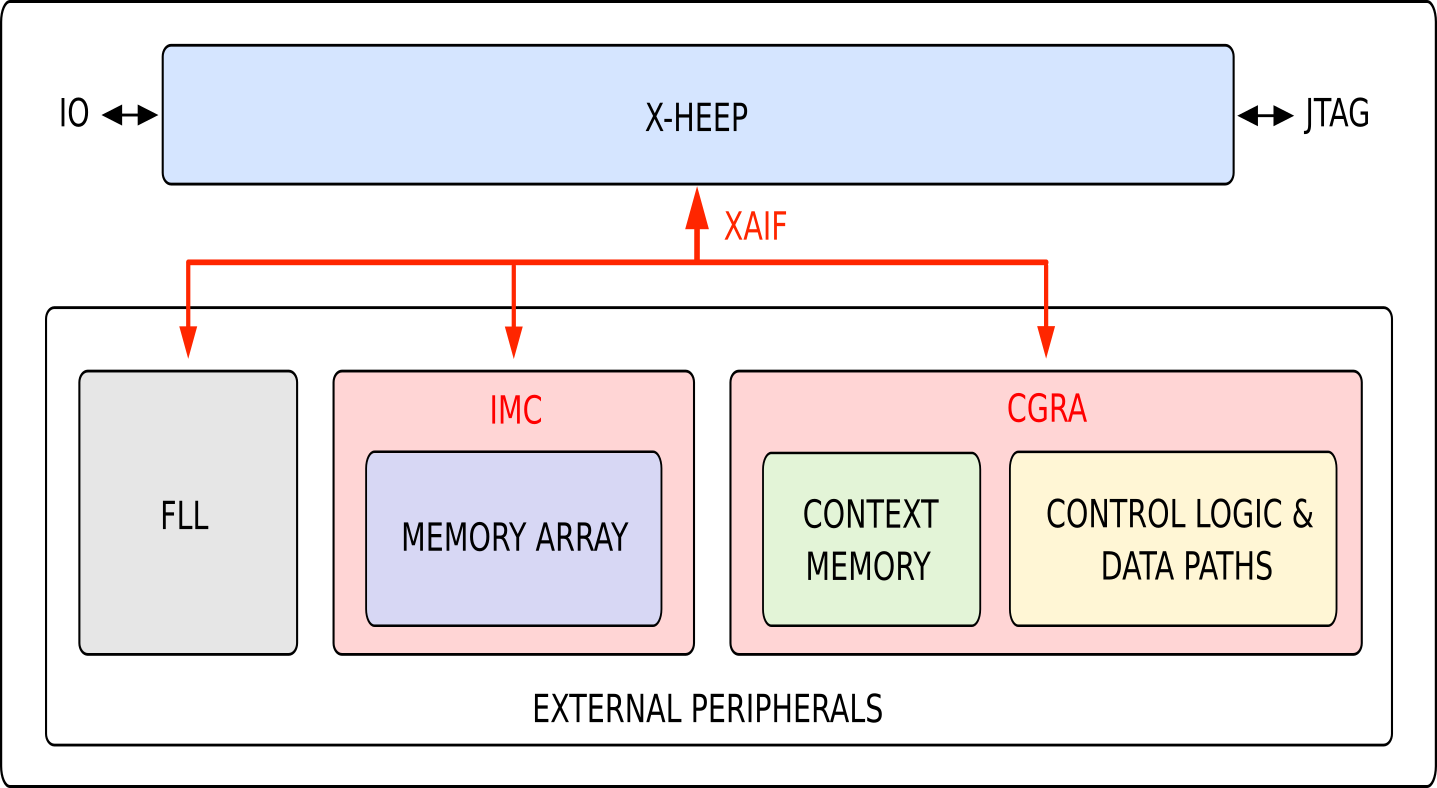}
                \vspace{10px}
                \caption{\textit{HEEPocrates} architecture. Power domains are visually marked using different colors. The components highlighted in grey are always on. The accelerator integration is highlighted in red.}
                \label{fig:heepocrates}
            \end{figure*}

        Figure \ref{fig:heepocrates} shows the \mbox{HEEPocrates} architecture highlighting how the CGRA and IMC accelerators are integrated to minimize power and maximize bandwidth.

        \subsubsection{X-HEEP configuration}
        
            We configured the \mbox{X-HEEP} host platform with~(1)~the CV32E20 core, which is optimal for running control tasks and offloading performance-intensive computations to the external accelerators while preserving low power consumption;~(2)~8 SRAM banks of \SI{32}{\kibi\byte} in contiguous addressing mode to accommodate variable lengths of data acquisitions while power-gating the unused banks on different applications;~(3)~a fully connected bus to provide high-bandwidth capabilities to the integrated accelerators;~(4)~all the available peripherals in place to deliver high flexibility;~(5)~a CGRA and IMC accelerators connected to the external XAIF interface.

        \subsubsection{CGRA accelerator~\cite{cgra}}
        
            This accelerator offers two slave ports, one to access the internal configuration registers, and one for the context memory, plus four master ports used to load and store data. The context memory stores the kernel's code executed by the four internal processing elements (PEs). Each PE is connected to a dedicated master port to read and write data from/to the \mbox{X-HEEP} main memory, independently. This allows a maximum bandwidth of \SI{128}{\bit} per bus cycle. To synchronize the CGRA and the CPU, the CGRA end-of-computation event is connected to the \mbox{X-HEEP} interrupt controller (PLIC) via the XAIF interface. 

            The CGRA is divided into two power domains: one for the control logic and the datapaths; and one for the context memory. The control logic and datapaths can be clock-gated or power-gated, while the context memory can be clock-gated, power-gated, or set in retention mode. This dual power domain structure enables clock-gating individual domains during short periods of inactivity and power-gating during extended non-use periods. Additionally, it offers the flexibility of independently setting the context memory in retention mode while clock-gating or power-gating the datapaths and control logic to save CGRA configuration time. The XAIF interface provides control over the various power modes, enabling the system to dynamically adjust its power consumption based on the operational requirements of the CGRA accelerator.

        \subsubsection{IMC accelerator~\cite{blade}}
        
            This accelerator offers one slave port to access its memory array. An internal controller decodes the memory requests and facilitates the transition of the accelerator between two modes: memory mode and computation mode. In memory mode, the memory space functions as a conventional memory bank. In contrast, the computation mode enables the execution of in-memory computations, eliminating the need for additional data transfers between the main memory and the accelerator.
        
            As for the CGRA, the IMC accelerator is placed in a separate power domain to save power when not used.

        \subsubsection{Frequency-locked loop~\cite{flleth}}

            We utilized the XAIF interface to connect the frequency-locked loop (FLL) responsible for generating the system clock from a \SI{32}{\kilo\hertz} external source. For real-time configurability, the FLL exposes a set of memory-mapped registers that enable the host CPU to adjust the system clock frequency during application execution, dynamically. This feature is precious during extended data acquisition periods in healthcare applications because it allows for reducing the system frequency to the minimum value required for acquiring the necessary biosignals, thereby minimizing dynamic power consumption. Lastly, the FLL can be also bypassed, allowing the external source to serve as the system clock.

    \subsection{FPGA implementation}

        We implemented \mbox{HEEPocrates} in FPGAs on the Zynq 7020, Zynq UltraScale+, and Artix 7 chips by Xilinx for early prototyping. This allows for the exploration of different \mbox{X-HEEP} configurations and accelerators to optimally tune the architecture for the healthcare domain.

    \subsection{Silicon implementation}

\begin{figure*}
                \centering
                \includegraphics[width=0.9\textwidth]{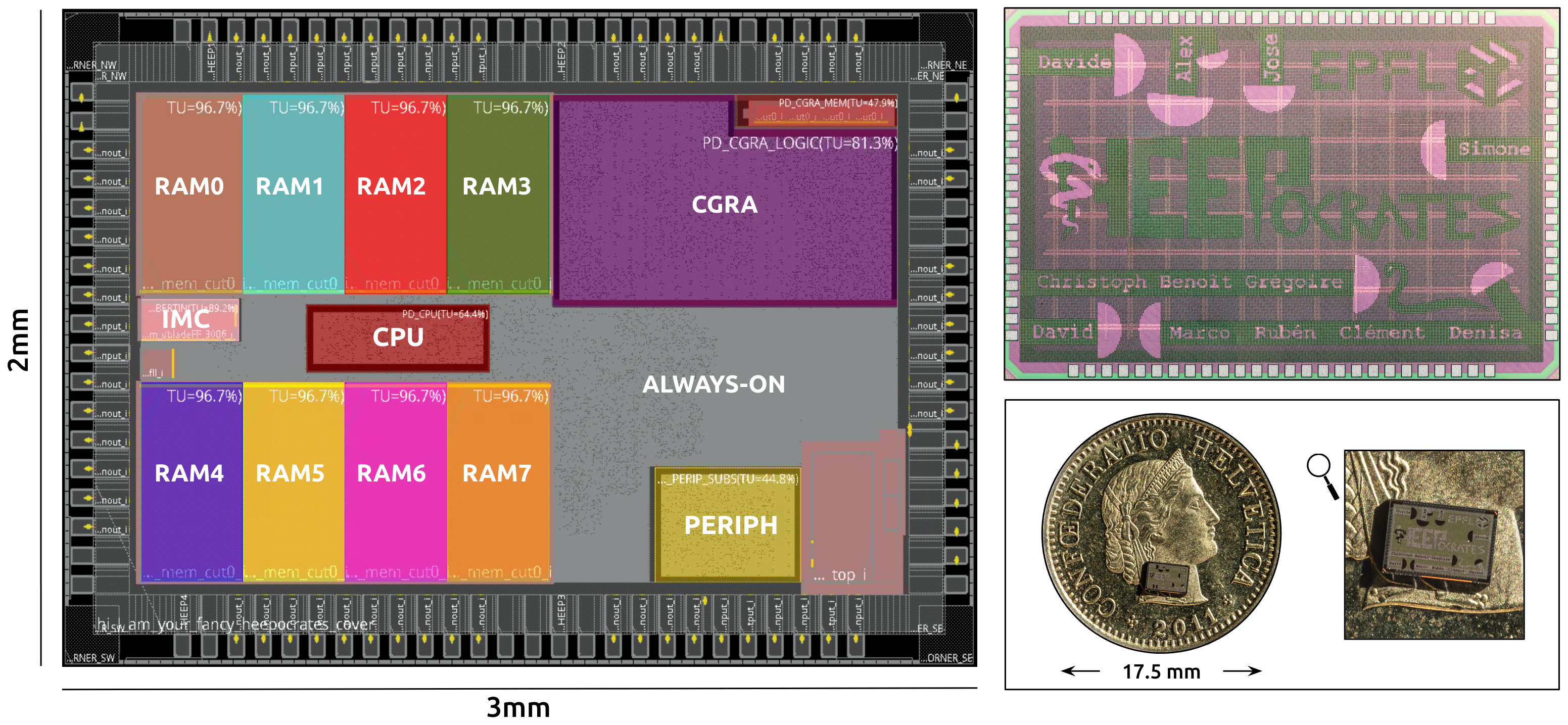}
                \vspace{5pt}
                \caption{\textit{HEEPocrates} layout, silicon photo, and physical chip (on a Swiss 5-cent franc coin).}
                \label{fig:layout}
            \end{figure*}

        After FPGA prototyping and exploration, we implemented \mbox{HEEPocrates} in silicon with TSMC \SI{65}{\nano\meter} low-power CMOS technology. Figure \ref{fig:layout} shows the \SI{6}{\milli\meter\squared} layout of \mbox{HEEPocrates}, with the power domains shown in different colors.
        
        For conducting our measurements, we developed a board specifically designed to accommodate our chip. \mbox{HEEPocrates} has been tested from \SI{0.8}{\volt} to \SI{1.2}{\volt}, achieving a maximum frequency of \SI{170}{\mega\hertz} and \SI{470}{\mega\hertz}, respectively. Power consumption ranges from \SI{270}{\micro\watt} at \SI{32}{\kilo\hertz} and \SI{0.8}{\volt}, to \SI{48}{\milli\watt} at \SI{470}{\mega\hertz} and \SI{1.2}{\volt}. Each phase of healthcare applications has been optimized to minimize power consumption.

        \subsubsection{Acquisition phase}

            Healthcare applications commonly feature an extended acquisition phase due to the low-bandwidth nature of biosignals and the typical lengthy data windows. During this phase, samples are gathered from external ADCs via SPI, or other I/O peripherals, and stored in memory by the main CPU or the DMA. We run this phase at \SI{1}{\mega\hertz}, \SI{0.8}{\volt} to minimize power while offering enough performance for the acquisition of bio-signals in the order of hundreds of Hertz. \mbox{HEEPocrates} consumes \SI{384}{\micro\watt} during acquisition when the complete system is active, and the host CPU is clock-gated when not used. However, power can be further optimized by switching off the unused memory banks, the peripheral domain, and the external accelerators for the entire acquisition period. This enables a reduction in power of \SI{19}{\percent}, which leads to \SI{310}{\micro\watt}. Furthermore, the CPU can be turned off during idle periods, i.e., when not used actively to acquire ADC samples, reaching the lowest power level of the system at \SI{1}{\mega\hertz} of \SI{286}{\micro\watt}, with a further reduction of \SI{8}{\percent}.

        \subsubsection{Processing phase}
        
            Upon completion of the acquisition phase, we run the processing phase at the maximum speed of \SI{170}{\mega\hertz}, \SI{0.8}{\volt} to minimize processing time and race to sleep. \mbox{HEEPocrates} consumes \SI{8.17}{\milli\watt} during the processing phase when the complete system is active and the CPU executes a matrix multiplication. Power can be further optimized \SI{6}{\percent} by turning off the unused memory banks, the peripheral domain, and the external accelerators, with a consumption of about \SI{7.68}{\milli\watt}. During the processing phase, the external accelerators can be individually powered on, and computationally intensive tasks can be offloaded by the main CPU to reduce the system's overall energy consumption. \mbox{HEEPocrates} consumes \SI{4.01}{\milli\watt} and \SI{1.65}{\milli\watt} when CNN algorithms are executed on the CGRA accelerator and IMC accelerator, respectively, at their maximum frequency of \SI{60}{\mega\hertz}. The host CPU, the unused memory banks, and the peripheral domain are powered off during accelerator activity.

%% file: sections/5-experimental-setup.tex
\section{Experimental setup}

    This section introduces a representative set of different families of microcontrollers commonly used in healthcare applications. Subsequently, it describes the biomedical applications that are included in our benchmark.

    \subsection{Healthcare microcontrollers} 
                
        Healthcare applications exhibit significant variability in acquisition and processing times, influenced by factors such as the length of sampling windows and the complexity of adopted algorithms. To address this variability, a diverse range of microcontrollers have been designed, each optimized to minimize power consumption during specific phases. The Apollo~3 Blue excels in acquisition phases, prioritizing power efficiency through its deep sleep mode, which ensures remarkably low power consumption when the system is inactive during idle periods. On the other hand, GAP9 takes the lead in processing phases thanks to its higher-performance core, which guarantees substantial reductions in processing time. The analysis of these two microcontrollers enables covering the entire spectrum of ultra-low-power edge devices, ranging from top-tier power efficiency, with Apollo~3 Blue, to top-tier performance, with GAP9. Furthermore, the frequent use of both microcontrollers in this domain demonstrates their capability to meet the rigorous demands of healthcare applications in terms of performance, power, and area. Table~\ref{fig:mcus} reports the features of the selected microcontrollers.
        
        \subsubsection{Apollo 3 Blue}

            This MCU is part of the Ambiq board and features an ARM Cortex-M4 core. The code is stored in the on-chip flash memory with zero overhead in instruction fetching, while the rest of the data resides either entirely in the SRAM when it fits or in both the SRAM and the flash. Unnecessary SRAM banks are turned off for the entire duration of the application. Its optimal processing configuration is \SI{0.7}{\volt}, \SI{48}{\mega\hertz}. However, we exploited the TurboSPOT mode to increase the frequency to \SI{96}{\mega\hertz} when required to meet the timing constraints of the benchmark applications. Moreover, during idle periods, the system enters its deep sleep mode, consuming approximately \SI{6}{\micro\ampere\per\mega\hertz}, where most of the system components are power-gated, with only a few power control modules active.

        \subsubsection{GAP9}

            This MCU is part of the GAP9EVK board and features one CV32E40P core, known as the fabric controller (FC), and a cluster (CL) with nine CV32E40P cores, which can be switched on and off. We execute the benchmark applications exclusively on the FC while power-gating the CL and unnecessary SRAM banks for the entire duration of the application. The application code and data are stored in the SRAM for maximum performance. Its optimal processing configuration is \SI{0.65}{\volt}, \SI{240}{\mega\hertz}. Furthermore, during idle periods, the system transitions into its sleep mode, where the majority of components are power-gated, except for memory banks, which enter a retention mode.

        \subsubsection{HEEPocrates}

            The application code and data are completely stored in the SRAM, when possible, or in a combination of the SRAM and the off-chip flash, connected through the SPI interface. The peripheral domain and the unused memory banks are also powered off throughout the entire duration of the application. We execute all the benchmark applications on the host CPU while power-gating the external accelerators. Moreover, we also accelerate CNN computations on the CGRA and IMC accelerators and showcase the energy improvement compared to running on the host CPU. We performed each measurement under the optimal operating conditions: \SI{170}{\mega\hertz} at \SI{0.8}{\volt}, for the host CPU; \SI{60}{\mega\hertz} at \SI{0.8}{\volt}, for the CGRA and IMC accelerators. During idle periods, the host CPU and the external accelerators are power-gated, and the system frequency is lowered to \SI{1}{\mega\hertz} to reduce power consumption.

            \captionsetup[figure]{name=Table}
            \setcounter{figure}{0}
            \begin{figure*}
                \caption{Microcontrollers commonly adopted in healthcare applications.}
                \vspace{10pt}
                \includegraphics[width=0.9\textwidth]{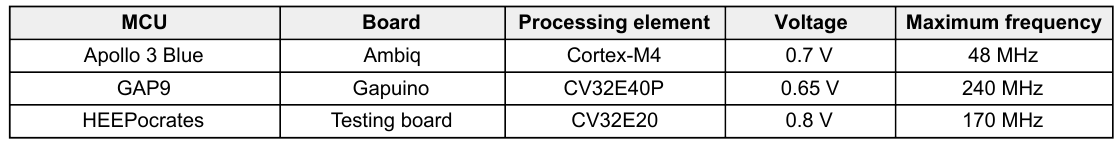}
                \label{fig:mcus}
            \end{figure*}
            \begin{figure*}
                \caption{Healthcare applications included in our benchmark.}
                \vspace{10pt}
                \includegraphics[width=0.9\textwidth]{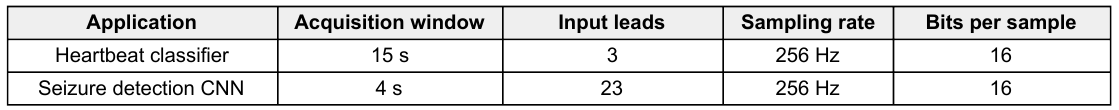}
                \label{fig:apps}
            \end{figure*}
            \setcounter{figure}{4}
            \captionsetup[figure]{name=Figure}

    \subsection{Healthcare applications}

        Table~\ref{fig:apps} reports the healthcare applications selected for our benchmark. Our selection ensures that we cover the full spectrum of ultra-low-power healthcare applications, ranging from acquisition-dominated, with the heartbeat classifier, to processing-dominated, with the seizure detection CNN. Moreover, these applications showcase computational algorithms of varying complexity, thereby enhancing the comprehensiveness of our analysis.

        \subsubsection{Heartbeat classifier~\cite{hcl}}
            
            This application is used to detect irregular beat patterns for common heart diseases through the analysis of electrocardiogram (ECG) signals. The most resource-intensive part of this application lies in the initial computation phase, specifically the morphological filtering, which consumes over \SI{80}{\percent} of the total processing time. Subsequently, the classification stage employs random projections. Initially, the algorithm processes a single input channel. If an abnormal heartbeat is detected, the analysis extends to the other leads for a more precise determination. Our testing scenarios involve input signals that all contain abnormal beats to evaluate the complete application pipeline. The input signal is derived from three distinct ECG leads, each sampled at \SI{256}{\hertz} with an accuracy of \SI{16}{\bit}. A \SI{15}{\second} acquisition window produces an input signal of \SI{22.5}{\kibi\byte}.

        \subsubsection{Seizure detection CNN~\cite{cnn}}

            This application is used to detect seizures in electroencephalography (EEG) signals. It features a CNN with three one-dimensional convolutional layers, each incorporating pooling and ReLU layers. \SI{90}{\percent} of the processing time is spent in convolutional computations, which mainly involve multiply and accumulate (MAC) and shift operations. Following each convolution, there is an overflow check and a maximum test for the pooling layer. Two fully connected layers end the network. The signal is sampled from 23 leads at a rate of \SI{256}{\hertz} with \SI{16}{\bit} accuracy and the acquisition phase lasts \SI{4}{\second}, resulting in an input signal size of \SI{46}{\kibi\byte}.

%% file: sections/6-experimental-results.tex
\section{Experimental results}

    In this section, first, we analyze the energy consumption of the proposed host platform, \mbox{HEEPocrates} (with the accelerators power-gated), in comparison with the selected state-of-the-art microcontrollers that may serve as host platforms. Subsequently, we assess the energy efficiency gained from leveraging the \mbox{HEEPocrates}' accelerators in comparison to execution on the host CPU.

    \subsection{Host platforms}

\begin{figure*}[t]
            \subfloat[Heartbeat Classifier.]{\includegraphics[width=2.9in]{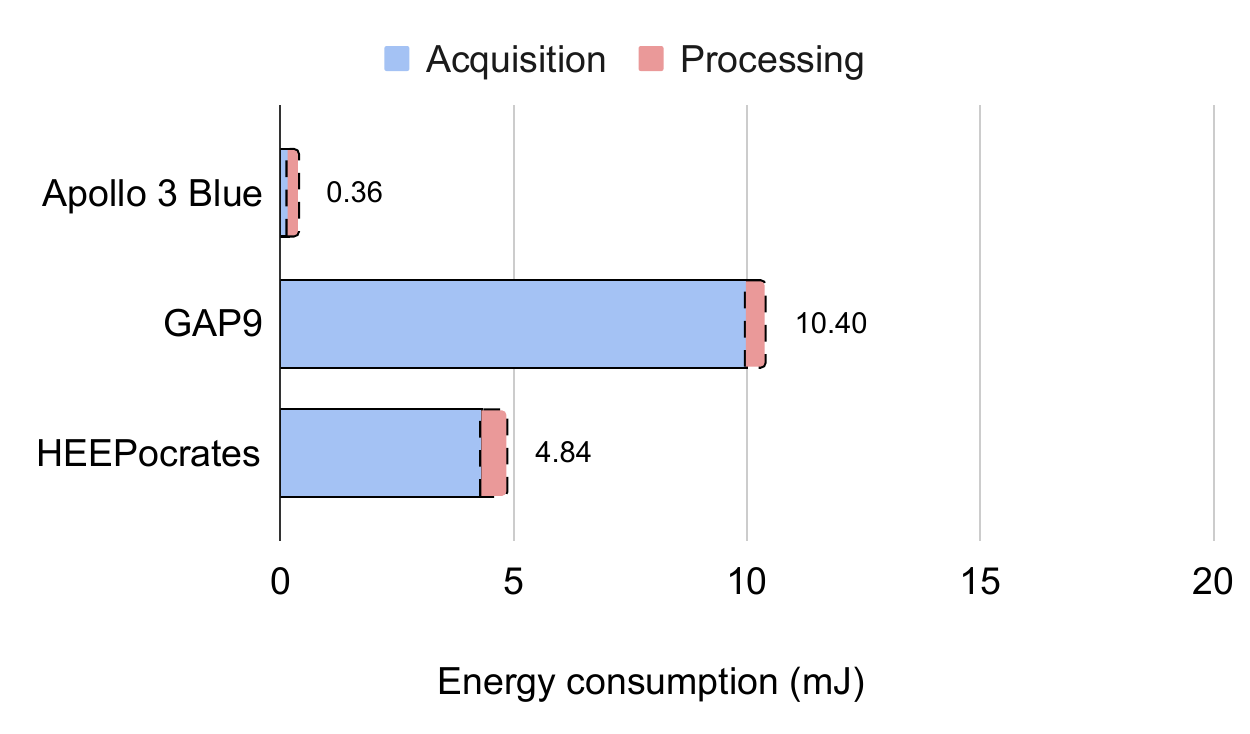}}
            \subfloat[Seizure Detection CNN.]{\includegraphics[width=2.9in]{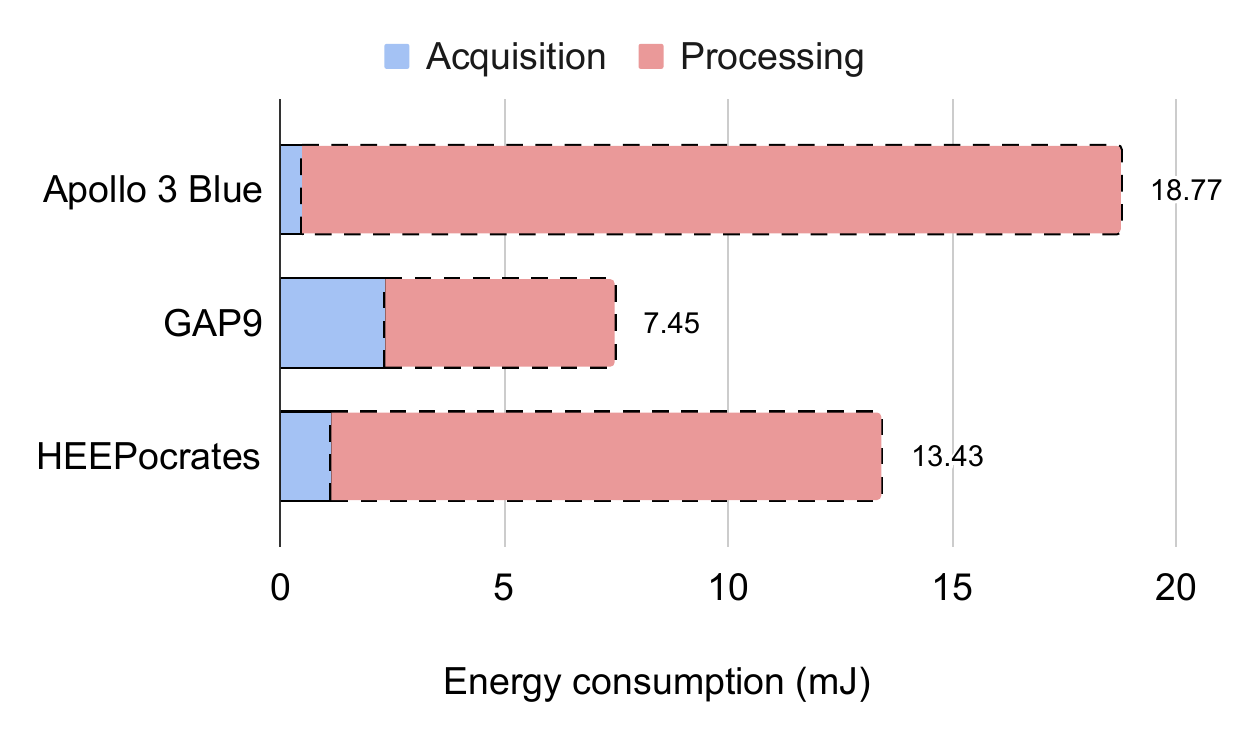}} 
            \vspace{5px}
            \caption{Energy consumption of our benchmark running on common healthcare microcontrollers and on \mbox{HEEPocrates} at \SI{0.8}{\volt}.}
            \label{fig:energy}
        \end{figure*}

        Figure \ref{fig:energy} illustrates the measured energy values for each healthcare application from our benchmark.
        
        The heartbeat classifier application exhibits an acquisition-driven nature, characterized by extended acquisition windows and a low sampling rate of \SI{256}{\hertz}. This forces microcontrollers to spend a significant amount of time in idle states during acquisition. In particular, the Apollo~3 Blue stands out for its energy efficiency, attributed to its remarkably low sleep mode of only \SI{6}{\micro\ampere/\mega\hertz}, where most of the system is power-gated, with only a few control modules active. On the contrary, GAP9 lacks aggressive sleep modes and keeps more modules always on, resulting in considerably higher energy consumption. Even during the processing phase, Apollo~3 Blue maintains a slight energy advantage over GAP9. This can be attributed to the optimized design of its CPU, the ARM \mbox{Cortex-M4}, which is more efficient for the specific operations required by this application, including logical and comparison operations, branches, as well as load and store instructions~\cite{hcl}.
        
        \mbox{HEEPocrates} positions itself in a middle ground during acquisition, offering a more robust sleep mode compared to GAP9. However, it does not reach the exceptionally low power consumption levels of Apollo~3 Blue due to the absence of aggressive sleep strategies for faster wake-up times, which includes in the always-on IPs more peripherals as an FLL, a pad controller, bus, a debug unit, and more general-purpose peripherals added for enhanced versatility (e.g. SPI, UART, etc.). 
        However, \mbox{HEEPocrates}' energy efficiency can be improved by removing the general-purpose peripherals, resulting in a \SI{27}{\percent} reduction in overall energy consumption. During processing, \mbox{HEEPocrates} consumes slightly higher energy compared to the other microcontrollers, due to its ultra-low-power CV32E20 core~\cite{zeroriscy} that is not optimized for performance like GAP9, and due to the higher-power consumption of the active part of the chip compared to Apollo-3, sitting \mbox{HEEPocrates} in the middle between the two. 
        
        The seizure detection CNN application is processing-dominated due to its computationally intense convolutional network, leading microcontrollers to spend the majority of their time in the processing phase. GAP9 emerges as the dominant contender in this phase, leveraging its high-performance core to achieve reduced processing times and efficient transitions to sleep. In contrast, the core of Apollo~3 Blue lacks sufficient computational power, resulting in an extended processing phase and increased energy consumption. However, during the acquisition phase, Apollo~3 Blue maintains dominance over GAP9 due to its more efficient sleep mode, resulting in lower energy consumption.
        
        \mbox{HEEPocrates} finds itself positioned between Apollo~3 Blue and GAP9 in both the processing and acquisition phases. During acquisition, it offers a more efficient low-power mode than GAP9 but does not reach the efficiency levels of Apollo~3 Blue, for the reasons explained earlier. In the processing phase, the higher performance of \mbox{HEEPocrates} allows for faster entry into the sleep state than Apollo~3 Blue but lags behind GAP9 due to its higher-frequency core. Notably, similar to the previous application, \mbox{HEEPocrates}' energy efficiency may be enhanced by removing general-purpose peripherals from the always-on domain, resulting in an overall energy reduction of about \SI{3}{\percent}.
        
        In conclusion, our analysis reveals the energy consumption alignment of \mbox{HEEPocrates} with state-of-the-art microcontrollers commonly adopted in healthcare applications. The performance and power efficiency of our platform falls between the top-tier power efficiency of Apollo~3 Blue and the top-tier performance of GAP9. This underscores that \mbox{HEEPocrates} achieves state-of-the-art energy efficiency figures across a wide range of real-world application profiles typical of the healthcare domain, ranging from acquisition-dominated to processing-dominated scenarios.

    \subsection{Accelerators}
    
\begin{figure}
        \includegraphics[width=0.5\textwidth]{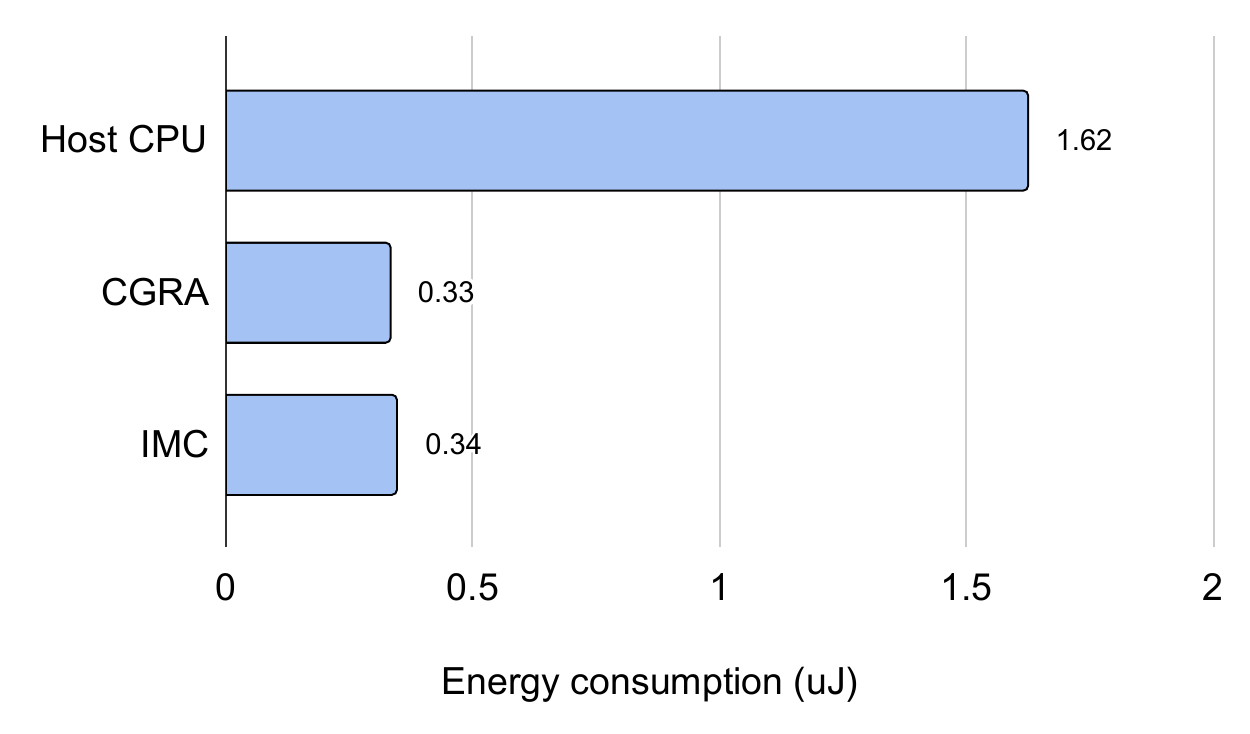}
        \vspace{5px}
        \caption{Energy consumption of \mbox{HEEPocrates} at \SI{0.8}{\volt} running a 16x16 convolution (3x3 filter) on the host CPU (at \SI{170}{\mega\hertz}) and the CGRA and IMC accelerators (at \SI{60}{\mega\hertz}).}
        \label{fig:accel}
    \end{figure}

        In Figure~\ref{fig:accel}, we compare the energy consumption of \mbox{HEEPocrates} while running a 16$\times$16 convolutional layer with a 3$\times$3 filter on the host CPU, the CGRA and IMC accelerators. Our results demonstrate an improvement in energy efficiency of approximately \SI{4.9}{\times} and \SI{4.8}{\times} achieved by exploiting the integrated CGRA accelerator and the IMC accelerator, respectively, compared to running on the host CPU. This improvement is attributed to the higher parallelism of the proposed accelerators, which compensates for the increased power consumption resulting from the more intense computation.
        
%% file: sections/7-conclusions.tex
\section{Conclusions}

    In this paper, we have explored the growth and increasing demand for efficient processing solutions in the field of edge computing, particularly in the context of new AI/ML applications. Persistent challenges arise from the limitations in performance and power consumption of edge devices, which impact overall energy efficiency.

    To address these challenges, heterogeneous architectures have emerged, presenting a promising solution by combining ultra-low-power host processors with specialized accelerators tailored to specific applications or domains.

    However, we have shown the limitations of existing host platforms in exploring the design space of accelerator-based ultra-low power edge AI platforms, as well as in providing the configurability and extendability options needed to integrate the large variety of custom accelerators and interfaces that exist nowadays. Consequently, extensive modifications to the RTL code are often required to integrate accelerators effectively, leading to high maintenance costs.

    To overcome these limitations, we introduced \mbox{X-HEEP}, an open-source solution designed specifically to support the integration and exploration of ultra-low-power edge AI/ML accelerators. The platform offers comprehensive customizability and extendability options via the proposed XAIF, which gathers all the requirements of state-of-the-art domain-specific solutions, as memory-mapped accelerators, including memory, processors, and peripherals with DMA-support, custom ISA co-processor, interrupts, and power saving strategies interface,  enabling designers to tailor the platform to meet the unique requirements of the target applications in performance, power, and area.

    \mbox{X-HEEP} provides configuration options to match specific application requirements by exploring various core types, bus topologies, and memory addressing modes. It also enables a fine-grained configuration of memory banks to match the constraints of the integrated accelerators. The platform prioritizes energy efficiency by implementing low-power strategies and integrating them with accelerators through dedicated power control interfaces. This cohesive integration ensures that all system components work together to maximize energy savings.

    To illustrate the practical benefits of \mbox{X-HEEP}, in this work, we presented a real-world integration example tailored for healthcare applications, which shows high variability among acquisition and processing-dominated application profiles. This example featured a CGRA accelerator and an IMC accelerator, both of which have proved to effectively reduce the overall energy consumption for this application domain. The resulting design, called \mbox{HEEPocrates}, has been implemented both in FPGAs on the Zynq 7020, Zynq UltraScale+, and Artix 7 chips by Xilinx, for early prototyping and exploration, and in silicon with TSMC \SI{65}{\nano\meter} low-power CMOS technology, for silicon validation. The fabricated chip can operate from \SI{0.8}{\volt} to \SI{1.2}{\volt}, achieving a maximum frequency of \SI{170}{\mega\hertz} and \SI{470}{\mega\hertz}, respectively. Its power consumption ranges from \SI{270}{\micro\watt} at \SI{32}{\kilo\hertz} and \SI{0.8}{\volt}, to \SI{48}{\milli\watt} at \SI{470}{\mega\hertz} and \SI{1.2}{\volt}.

    To measure the performance and versatility of the proposed design, we analyze the execution of an illustrative real-life set of edge AI/ML benchmarks that combines ultra-low power healthcare applications from the latest advances in the field, showing high variability in the execution profile. Through the execution of our benchmark and the measurement of the energy consumption of the chip, we demonstrated \mbox{HEEPocrates'} alignment with other state-of-the-art microcontrollers that are frequently employed in healthcare applications. This is achieved thanks to a balanced trade-off between fine-grain power domains, to reduce power consumption during acquisition phases, and on-demand accelerator capabilities, to speed up the execution of processing phases, resulting in a good trade-off between acquisition-dominated and processing-dominated applications. 
    These results also showcase the representativeness of the experiments that other researchers could perform after integrating their accelerators with \mbox{X-HEEP}. Lastly, we proved the energy benefit of \SI{4.9}{\times} and \SI{4.8}{\times} gained by exploiting the integrated CGRA accelerator and IMC accelerator, respectively, compared to running on the host CPU.

    In conclusion, the introduction of the \mbox{X-HEEP} platform leads to a significant step forward in overcoming the challenges faced in the field of edge computing. By providing extensive options for customizability and extendability, prioritizing energy efficiency, and presenting a practical real-world integration example, \mbox{X-HEEP} presents itself as an innovative platform, empowering designers and researchers to create efficient heterogeneous edge AI/ML computing systems.

\section{Acknowledgements}

    We would like to thank the entire \mbox{X-HEEP} team for their great contribution to the platform.